\documentstyle[12pt]{article}
\newcommand{\beq}{\begin{equation}}
\newcommand{\eeq}{\end{equation}}
\newcommand{\beqa}{\begin{eqnarray}}
\newcommand{\eeqa}{\end{eqnarray}}
\newcommand{\ba}{\begin{array}}
\newcommand{\ea}{\end{array}}

\begin{document}

\begin{flushright}
Preprint CAMTP/98-2\\
April 1998\\
\end{flushright}

\parindent 0pt
\vskip 0.5 truecm
\begin{center}
\Large
{\bf  On the Green function of linear evolution equations 
for a region with a boundary}\\
\vspace{0.25in}
\normalsize
George Krylov$^{\dag \, \ddag}$ and Marko Robnik$^{\dag}$
\footnote{e--mails:
george.krylov@uni-mb.si, robnik@uni-mb.si}\\

\vspace{0.3in}
\dag Center for Applied Mathematics and Theoretical Physics,\\
University of Maribor, Krekova 2, SI--2000 Maribor, Slovenia\\
\ddag Department of Physics, Belarusian State University,\\
Fr. Skariny av. 4, 220050 Minsk, Belarus\\

\end{center}
\parskip 10pt

\vspace{0.3in}

\normalsize
\noindent
{\bf Abstract.}  We derive a closed form expression for the
Green function of linear evolution equation with the Dirichlet 
boundary condition for an arbitrary region, based on 
singular perturbation approach to boundary problems.

\vspace{0.6in}

Published in {\bf Journal of Physics A: Mathematical and General}\\
Vol. 32 (1999) 1261-1267
\normalsize
\vspace{0.1in}

\newpage

\section{Introduction}
The boundary value problem for linear operators in non-trivial regions leads to
complications and quite often it is necessary to resort
to the numerical analysis even for the simplest
operators which possess a known kernel in the whole space, such as e.g. 
Laplace operator in the $\mbox{\bf R}^n$ \cite{Morse}.

The possibility of a new approach to these problems appeared 
along with developing 
the theory of point interactions in quantum mechanics, firstly stimulated
by the famous Kronig-Penney model \cite{C-P}, and systematically investigated
in \cite{Albeverio} where self-adjoint extensions for Hamiltonian with
point-like interactions have been constructed  so that the 
explicit form for
resolvent has been obtained for some physically 
significant systems. It is in reference
\cite{Albeverio} that it has been already pointed out that 
the limiting case of
infinitely strong point interaction allows 
effectively to split the space in two
separated regions that lead to two boundary problems on half line.

This important trick has been successfully developed in 
\cite{Grosche1,Grosche2,Grosche3} 
and it allows one to write down the explicit expressions for the 
Green function of
Schr\"odinger equation for the particle in one dimensional and radial boxes
with Dirichlet and Neumann boundary conditions provided the appropriate
problem in the whole space has been solved.   
The technique that has
been successfully used
for such a derivation is a direct summation of perturbation series
(Dyson series \cite{Kleinert}), 
effectively
leading to geometrical progression due to the
specific form of the perturbation.
We would like to stress here, that numerous analytical results obtained
up to now are related to quantum one-dimensional problems (as e.g., Krein's
formula \cite{Albeverio}), 
effective one-dimensional problems after variables' separation
('radial problem'), and
point-like interactions (see \cite{Albeverio,Jakiw} and 
recent book \cite{G-S} for 
detailed references which include both solvable 
$\delta$-perturbation cases and boundary value problems).

The question naturally arising here is whether it is possible to
generalize these constructions to higher dimensional boundary problems.
As we will see, and this will be the main aim of this work,
it can be done at least for Dirichlet boundary conditions
for arbitrary linear evolution operators in topologically trivial region
(homeomorphic to a  ball in $R^d$), 
when assuming some natural
conditions for the propagator of ``free particle'' to be fulfilled.
It is worthwhile to mention here that we intend 
neither to construct the
self-ajoint extension of appropriate singular perturbed operators
\cite{Albeverio}, nor to perform a detailed investigation of
convergence properties of appropriate perturbation series
(as is well known the answer may be negative even for regular
perturbations, see e.g. \cite{Kleinert}). To the 
contrary, we intend to propose explicit
construction of Green function for non-separable case and demonstrate its
validity  by known examples. Generally speaking the problem
we will solve can be
formulated on an abstract level of the theory of linear operators by
introducing some generalization of projector operators, but we
expect that it could only shadow simple foundation of the approach we
use. Moreover, although, admittedly, the manipulations with  perturbation series
and the subsequent limit of infinitely large coupling constant are
rather formal,
we do not know another way to obtain the Green function
representation for general boundary problem that we will construct in this paper.

\section{Series summation for singular perturbed system}
Let us have a linear evolution equation in $R^d$ of the form
\begin{equation}  \label{eq:evol1}
 \left(
   \frac{\partial}{\partial t} - \hat {\cal L} \right)\Psi({\bf{x}})=0, 
\end{equation}
with explicitly time independent operator $\hat {\cal L}$ acting on 
the function defined over ${\rm{R}^d}$ and obeying
Dirichlet boundary condition $\Psi ({\bf{x}})|_\Gamma=0$, where 
$\Gamma= \{ {\bf{x}}: P({\bf{x}})=0 \} $ is the boundary (hyper-surface)
of the region ${\cal B}$ under consideration, in which we seek
the solutions of equation (\ref{eq:evol1}).

We start from a consideration of the ``free particle'', omitting the boundary
condition, and assuming that propagator for that case, given by
\begin{equation}\label{eq:prop00}
 K^0({\bf{x}}',{\bf{ x}}''; t)=
<{\bf{ x}}''| e^{ \hat {\cal L} t}|{\bf{ x}}'>\theta(t),
\end{equation}
is already known.
Here $\theta$ is the Heaviside unit-step function, incorporated into
(\ref{eq:prop00}) to ensure causality property 
$ K({\bf{x}}',{\bf{ x}}''; t)=0$, when $t<0$.
The propagator 
$K$ possesses the composition  property 
\begin{equation}\label{eq:sg}
 K({\bf{x}}',{\bf{ x}}''; t)=
 \int\limits_{\rm{R}^d} d{\bf{x}}_1
 K({\bf{x}}',{\bf{ x}}_1; t_1)
 K({\bf{x}}_1,{\bf{ x}}''; t-t_1).
\end{equation}
which follows from the semi-group property with respect 
to the time evolution, which in turn is automatically fulfilled
for an explicitly time-independent operator $\hat {\cal L}$ and
the decomposition of unity in an appropriate functional space, namely
\begin{equation}  \label{eq:unity}
  \mbox{ Id}=\sum_x \left|x\right>\left<x\right|,
\end{equation}
where summation (integration) is performed over discrete 
(continuous) index
enumerating states (see e.g, \cite{Kleinert}). 
These properties are natural for most of physically significant
models so that our consideration is  not very restrictive.

Now we will emulate the boundary condition
by introducing into (\ref{eq:evol1}) the
additional singular potential term of the form 
$V^\delta=-\gamma \delta_P({\bf{x}})$. 
The generalized $\delta$-function used here is a distribution 
concentrated on the
hyper-surface $P$ \cite{Shilov}.  
In the limiting case 
of $\gamma \to \infty$
the corresponding one dimensional problem turns out to be a 
Dirichlet boundary problem on the half line \cite{Albeverio} 
(see the discussion in introduction). We will demonstrate
explicitly that the same situation is met in higher dimensions.

We will use the method expounded in 
\cite{Albeverio,Grosche1,Grosche2,Grosche3,Grosche4}, performing 
a perturbation expansion, starting from
the formula for the propagator of the singular perturbed problem
\begin{equation}\label{eq:prop0}
 K({\bf{x}}',{\bf{ x}}''; t)=
<{\bf{ x}}''| e^{ (\hat {\cal L}+V^\delta) t}|{\bf{ x}}'>\theta(t).
\end{equation}

The formal perturbation series over powers of $V^\delta$ can be constructed
as in the quantum mechanics \cite{F-H} and reads
\begin{eqnarray}\label{eq:series0} \nonumber
 K^\delta ({\bf{x}}',{\bf{ x}}''; t)=
 K^0({\bf{x}}',{\bf{ x}}''; t)+ \sum\limits_{n=1}^{\infty}\gamma^n
 \int\limits_{0}^{t} dt_1 \int\limits_{\rm{R}^d} d {\bf x}_1
 K^0({\bf{x}}',{\bf{ x}}_1; t_1-0)\delta_P({\bf{x}}_1) \times \\ 
 \prod\limits_{j=2}^{n} \left[\; \int\limits_{0}^{t} dt_j 
 \int\limits_{\rm{R}^d} d {\bf x}_j
 K^0({\bf{x}}_{j-1},{\bf{ x}}_{j}; t_{j}-t_{j-1}) \delta_P({\bf{x}}_j)\right]
 K^0({\bf{x}}_n,{\bf{ x}}''; t-t_n).
\end{eqnarray} 
The convergence questions appearing at this moment should be treated 
for every problem considered separately, e.g.  for Schr\"odinger equation
the existence of well defined Green function has been proven rigorously
in some cases \cite{Albeverio}. For arbitrary linear evolution equation
we must stay on the formal level only to go further.

After performing the Laplace transformation for the Green function,
defined by
\begin{equation}\label{eq:laplas}
 G({\bf{x}}',{\bf{ x}}''; E)= \int\limits_{0}^{\infty} e^{-Et}
 K({\bf{x}}',{\bf{ x}}''; t) dt,
\end{equation}
the following series representation can be written
\begin{eqnarray}\label{eq:green1} \nonumber
 G^\delta ({\bf{x}}',{\bf{ x}}''; E)=
 G^0({\bf{x}}',{\bf{ x}}''; E)+ \sum\limits_{n=1}^{\infty}\gamma^n
 \int\limits_{\rm{R}^d} d {\bf x}_1
 G^0({\bf{x}}',{\bf{ x}}_1; E)\delta_P({\bf{x}}_1) \times 
\\ 
 \prod\limits_{j=2}^{n} \left[\;  
 \int\limits_{\rm{R}^d} d {\bf x}_j
 G^0({\bf{x}}_{j-1},{\bf{ x}}_{j}; E) \delta_P({\bf{x}}_j)\right]
 G^0({\bf{x}}_n,{\bf{ x}}''; E).
\end{eqnarray} 

The behaviour of $ G^0({\bf{x}}',{\bf{ x}}''; E) $ 
at coincident space
arguments in spaces with $d>2$ may lead to divergence of integrals in 
(\ref{eq:green1}),
but we will not discuss this in details, since the appropriate 
procedures of regularizations are
well known (for point-like perturbations in quantum mechanics see e.g., 
\cite{Albeverio,Grosche4}). We only point out that for most interesting cases
of two and three-dimensional quantum problems there are no singularities
within our approach,
opposite to the models described in \cite{Grosche4}. Indeed, the  short
distance behaviour of the Green function in d-dimensional spaces is
(\cite{G-S}, f.6.2.1.2)
\begin{equation}\label{free}
 G(\mbox{\bf x}',\mbox{\bf x}'',k) \propto |\mbox{\bf x}'-{\bf x}''|^{1-d/2} 
Y_{1-d/2}(k|{\bf x}'-{\bf x}''|),
\end{equation}
where $Y_n(x)$ is Bessel function \cite{Watson},
so that the relevant underlying 
singularities are integrable for d=2,3.
For higher dimension (or) and other operator ${\cal L}$, 
some sort of regularization should be used as e.g. in
\cite{Grosche4}.

Returning to our problem, now  
we can introduce new coordinates by the map
$F: {\bf{x}}=\{x_1,..x_d\} \mapsto {\bf{y}}=\{y_1,..,y_d \}$ with
Jacobian  $J=\frac{(y_1,..,y_d)}{(x_1,..,x_d)}$ so that the
equation of hyper-surface $P$ will be given by $ y_d=\eta$ 
(see e.g.,\cite{Shilov}).
We designate all coordinates except the last one namely 
$\{y_i : i=1,..d-1 \}$, by $\Omega$, so that
${\bf{y}}=\{\Omega,y_d\}$. Then, the integrations over $\delta$-functions
are simply projections on the submanifold, defined by $y_d=\eta$ and we get
\begin{eqnarray}\label{eq:green2} \nonumber
 G^\delta ({\bf{x}}',{\bf{ x}}''; E)=
 G^0({\bf{x}}',{\bf{ x}}''; E)+ \sum\limits_{n=1}^{\infty}\gamma^n
 \int\limits_{P} \sqrt{g}_1 d {\Omega}_1
 G^0(\Omega',(y_d)',\Omega_1,\eta; E) \times \\ 
 \prod\limits_{j=2}^{n} \left[\;  
 \int\limits_{P} \sqrt{g}_j d {\Omega}_j
 G^0(\Omega_{j-1},\eta,\Omega_{j},\eta; E)\right]
 G^0(\Omega_n,\eta,\Omega'',(y_d)''; E),
\end{eqnarray} 
where the integration is performed over the hyper-surface $P$,
$g={\rm{det}}(g^{\mu\nu})$, 
$g^{\mu\nu}=\frac{\partial x^\mu}{\partial y^\nu}|_P$ 
is an induced metric tensor on 
 $P$ and we introduce coordinates $\Omega, y_d$ corresponding to
the initial and final points ${\bf{x}}',{\bf{x}}''$.
Now we want to expand the Green function 
$ G^0(\Omega_{j-1},\eta,\Omega_{j},\eta; E)$
in a series of functions defined on $P$. 
Let us choose  an appropriate
full (complete) orthonormal system of functions 
 $\{f_\kappa(\Omega)\}$ on the boundary $\Gamma$ of the region ${\cal B}$, 
where $\kappa$ is some multi-index enumerating the system $f$, 
with a standard $L_2(\Omega)$ scalar product 
\begin{equation}\label{scalar-prd}
 <f_\kappa,f_{\kappa'}>=\int\limits_{\Omega} \sqrt{g(\Omega)} 
  \bar f_\kappa(\Omega) f_{\kappa'}(\Omega) d\Omega=\delta_{\kappa,\kappa'}. 
\end{equation}
For example, the case of axially symmetric closed surfaces has been
recently treated by Prodan \cite{Prodan}, where the projection of the
resolvent operator on such surfaces has been investigated.

We represent $G^0$ in the form
\begin{equation}\label{eq:series}
  G^0(\Omega',\xi,\Omega'',\eta; E) =
  \sum\limits_{\kappa',\kappa''}{\cal G}_{\kappa',\kappa''}(\xi,\eta;E)
   f_{\kappa'}(\Omega')\bar f_{\kappa''}(\Omega'')
\end{equation}
so that the coefficient ${\cal G}_{\kappa',\kappa''}(\xi,\eta;E)$ is 
expressed as 
\begin{equation}\label{eq:coeff}
{\cal G}_{\kappa',\kappa''}(\xi,\eta;E)= 
\int\sqrt{g(\Omega')g(\Omega'')} \, G^0(\Omega',\xi,\Omega'',\eta; E)
   f_{\kappa'}(\Omega')\bar f_{\kappa''}(\Omega'')\,  d\Omega'  d\Omega''
\end{equation}

Then, substituting (\ref{eq:series}) into (\ref{eq:green2})  we get
\begin{eqnarray}\label{eq:green3} \nonumber
 G^\delta ({\bf{x}}',{\bf{ x}}''; E)=
 G^0({\bf{x}}',{\bf{ x}}''; E)+ \sum\limits_{n=1}^{\infty}\gamma^n
 \int\limits_{P} \sqrt{g}_1 d {\Omega}_1 
  \sum\limits_{\kappa',\kappa_1}{\cal G}_{\kappa',\kappa_1}((y_d)',\eta;E) 
\times \\ \nonumber
   f_{\kappa'}(\Omega')\bar f_{\kappa_1}(\Omega_1)
 \prod\limits_{j=2}^{n} \left[\;  
 \int\limits_{P} \sqrt{g}_j d {\Omega}_j
  \sum\limits_{\kappa_{j-1},\kappa_j}{\cal G}_{\kappa_{j-1},\kappa_j}(\eta,\eta;E)
   f_{\kappa_{j-1}}(\Omega_{j-1})\bar f_{\kappa_j}(\Omega) \right]
\times \\ \nonumber
  \sum\limits_{\kappa_n,\kappa''}{\cal G}_{\kappa_n,\kappa''}((y_d)'',\eta;E)
   f_{\kappa_n}(\Omega_n)\bar f_{\kappa''}(\Omega'') = \\ \nonumber
 G^0({\bf{x}}',{\bf{ x}}''; E)+
 \gamma \sum\limits_{\kappa',\kappa_1,\kappa_n\kappa''} 
 {\cal G}_{\kappa',\kappa_1}((y_d)',\eta;E) 
 {\cal G}_{\kappa_n,\kappa''}(\eta,(y_d)'';E)
 f_{\kappa'}(\Omega')\bar f_{\kappa''}(\Omega'') \times \\ 
 \left[ \delta_{\kappa_1,\kappa_n}+ 
 \gamma{\cal G}_{\kappa_1,\kappa_n}(\eta,\eta;E)+
\gamma^2\sum\limits_{\kappa_2}
 {\cal G}_{\kappa_1,\kappa_2}(\eta,\eta;E)
{\cal G}_{\kappa_2,\kappa_n}(\eta,\eta;E)
 + \dots \right]  
\end{eqnarray} 
where we used the orthonormality of the functions (\ref{scalar-prd}). 
After summing up the geometrical progression  we obtain
\begin{eqnarray}\label{eq:green4} \nonumber
 G^\delta ({\bf{x}}',{\bf{ x}}''; E)=
 G^0({\bf{x}}',{\bf{ x}}''; E)+ \\ 
  \sum\limits_{\kappa',\kappa''}
 \left[
  {\cal G}((y_d)',\eta;E)(\gamma^{-1}-{\cal G}(\eta,\eta;E))^{-1}
  {\cal G}(\eta,(y_d)'';E) 
 \right]_{\kappa',\kappa''}
   f_{\kappa'}(\Omega')\bar f_{\kappa''}(\Omega'').
\end{eqnarray} 
For the brevity of notation we used matrix form within the square brackets,
and $(\gamma{\cal G})^n$ is an ordinary matrix power.
Taking the limit $\gamma\to \infty$ we get finally
\begin{eqnarray}\label{eq:green5} \nonumber
 G^\delta ({\bf{x}}',{\bf{ x}}''; E)=
 G^0({\bf{x}}',{\bf{ x}}''; E)-  \\
  \sum\limits_{\kappa',\kappa''} 
 \left[
  {\cal G}((y_d)',\eta;E){\cal G}(\eta,\eta;E)^{-1}
  {\cal G}(\eta,(y_d)'';E) 
 \right]_{\kappa',\kappa''}
   f_{\kappa'}(\Omega')\bar f_{\kappa''}(\Omega''),
\end{eqnarray} 
which is the main result of our paper.
\section{Discussion}
As it is easily  seen, the last formula solves an appropriate
Dirichlet boundary problem. Indeed, the statement that the 
object constructed above
satisfies the differential equation (\ref{eq:evol1}) is evident from 
the construction and, if $(y_d)'=\eta $ or $(y_d)''=\eta $,
that is if the initial or the final points are on a boundary, the term in 
the square brackets simply gives
the free Green function expansion coefficient and the whole sum becomes 
the free Green
function cancelling the first term in (\ref{eq:green5}), which means that
the Dirichlet boundary condition is obeyed.

It is worthwhile to point out here that the spectrum of the system 
under consideretion is given by such values of 
$E$ that ${\cal G}(\eta,\eta;E)$ is
non-invertible. In two dimensional case the multi-index $\kappa$ becomes
an ordinary one and we obtain the
condition  of vanishing determinant
\begin{equation}
\mbox{Det}\; \; {\cal G}(\eta,\eta;E)=0.
\end{equation}
 From the last equation it is easily seen that our approach looks like
some alternative and generalization of the
boundary integral method \cite{BIM, BerW}, where the spectrum of
2-D  billiard
can be obtained based on an integral of Green function's 
normal derivative
over the boundary.

It is also easy to demonstrate that formula (\ref{eq:green5}) 
leads to a known one for the case of the separability of variables.
Let us do it explicitly for the 2-D case of quantum
particle in a circular
region ($\Gamma=\{{\bf x}: |{\bf x}|=R\}$). 
An appropriate formula for the Green function in polar
coordinates ($r,\phi$) reads \cite{Grosche1} (see also \cite{our1}
for alternative derivation)
\begin{eqnarray}\label{Gr}
 G ({\bf{x}},{\bf{ x}}'; E)=
 \sum\limits_{m=-\infty}^{\infty} G_l(r,r',E) e^{im(\phi'-\phi)},
  \\
\label{Gr1}
  G_l(r,r';E)=G^{0}_l(r,r';E)-
 \frac{G^{0}_l(r',R;E)G^{0}_l(R,r';E)}{G^{0}_l(R,R;E)}.
\end{eqnarray}

A natural choice of the functions' family is of course 
$f_m(\phi)=\exp\{im\phi\}$. Expanding the free particle Green function
in the same manner as in (\ref{Gr})  
and calculating the coefficient of 
equation (\ref{eq:series}) one can see that
\begin{equation}\label{gmm}
{\cal G}_{mm'} (r,r'; E)=G^0_m (r,r';E)\delta_{mm'},
\end{equation}
so that the matrix inversion becomes trivial and after the substitution
of (\ref{gmm}) into (\ref{eq:green5}), 
formula (\ref{Gr1}) follows immediately. Similar
arguments may be used for other separable quantum problems.

Thus, we
see, that indeed  we have successfully 
 constructed the explicit representation
for the Green function of linear evolutional
equation with Dirichlet boundary condition, based on
Green function in the whole space, thereby
generalizing results already known in
separable cases in quantum mechanics.

The formula (\ref{eq:green5}) can be rewritten in a more formal way, 
introducing
the series expansion of $ G^\delta ({\bf{x}}',{\bf{ x}}''; E)$ 
in a manner like
(\ref{eq:series}) for $G^0$. Then,
\begin{eqnarray}\label{eq:coeff1} \nonumber
{\cal G}^\delta_{\kappa'\kappa''} ((y_d)',(y_d)''; E)=
 {\cal G}_{\kappa'\kappa''} ((y_d)',(y_d)''; E)-  \\ 
\left[
  {\cal G}((y_d)',\eta;E){\cal G}(\eta,\eta;E)^{-1}
  {\cal G}(\eta,(y_d)'';E) 
 \right]_{\kappa',\kappa''}
\end{eqnarray} 
or in operator notation
\begin{eqnarray}\label{eq:oper}\nonumber
\hat {\cal G}^\delta ((y_d)',(y_d)''; E)=
 \hat {\cal G}((y_d)',(y_d)''; E)- \\  
  \hat {\cal G}((y_d)',\eta;E)\hat{\cal G}(\eta,\eta;E)^{-1}
  \hat{\cal G}(\eta,(y_d)'';E).
\end{eqnarray} 
The last expression is suitable for further formal manipulations in the
case of double $\delta$-perturbation  
$V=\gamma (\delta(y_d-a) +\delta(y_d-b))$, 
where the system is being "squeezed" into a narrow shell
$a\le \eta\le b$, simulating the quantization on a
hyper-surface in the limit $a\to b$ in a manner similar to
\cite{Grosche1} (see the equation~2.15 in \cite{Grosche1}), 
but the detailed analysis 
will be published elsewhere.

It should be mentioned also, that our approach can be modified
also to be used
for more  general perturbation of the form 
$\tilde V^\delta=- \gamma h({\bf{x}}) \delta_P({\bf{x}})$ with 
arbitrary function $h$. 
Then similar arguments show that the only difference from the
case  considered above  is 
that one should change $\sqrt{g}$ to $\sqrt{g}\;h(\Omega, \eta)$. Then, we
should use another function family for the expansion or, it may be more
convenient to expand into the series not the Green function but the product
\begin{equation}\label{eq:series1}
  \sqrt{h(\Omega; \xi)h(\Omega'', \eta)} G^0(\Omega',\xi,\Omega'',\eta; E) =
  \sum\limits_{\kappa',\kappa''}\tilde {\cal G}_{\kappa',\kappa''}(\xi,\eta;E)
   f_{\kappa'}(\Omega')\bar f_{\kappa''}(\Omega''),
\end{equation}
and the final formula becomes
\begin{eqnarray}\label{eq:green6} \nonumber
 \tilde G^\delta ({\bf{x}}',{\bf{ x}}''; E)=
 G^0({\bf{x}}',{\bf{ x}}''; E)-
  \sum\limits_{\kappa',\kappa''} 
   \frac{f_{\kappa'}(\Omega')\bar f_{\kappa''}(\Omega'')}
    {\sqrt{h(\Omega'', (y_d)')h(\Omega'', (y_d)'')}} \times \\
 \left[
  \tilde {\cal G}((y_d)',\eta;E)\tilde{\cal G}(\eta,\eta;E)^{-1}
  \tilde{\cal G}(\eta,(y_d)'';E) 
 \right]_{\kappa',\kappa''}.
\end{eqnarray} 

\noindent
Such a generalization may be useful for the systems with the boundary whose
initial shape is not very convenient for the construction of the function family
set $f_\kappa$ and when it is easy to perform some transformations
before using the proposed approach. In this case, after a transformation to
the new coordinates 
(and, e.g., accompanied by the ``local time rescaling'' \cite{Leschke1}),
the initially pure $\delta$-function perturbation really transforms to a
non-uniform one like discussed above.

\section*{Acknowledgements}
\par
The financial support by the Ministry of Science
and Technology of the Republic of Slovenia is acknowledged with
thanks.  This work was supported also by the Rector's Fund of the
University  of Maribor. G.K. acknowledges the support by
the postdoctoral  Grant No. SZF-CAMTP01-MZS/1997
of the  Slovenian Science Foundation.

\end{document}